\documentclass[prl,twocolumn,superscriptaddress]{revtex4}

\usepackage{graphicx}
\usepackage{dcolumn} 
\usepackage{amsmath} 
\usepackage{color} 

\begin{document}

\title{Interplay between electron correlations, magnetic state, and structural confinement in LaNiO$_3$ ultrathin films}

\author{N.~O. Vambold}
\affiliation{Institute of Physics and Technology, Ural Federal University, 620002 Yekaterinburg, Russia}
\email{nikitavamb@gmail.com}

\author{G.~A. Sazhaev}
\affiliation{Institute of Natural Sciences and Mathematics, Ural Federal University, 620002 Yekaterinburg, Russia}

\author{I.~V. Leonov}
\affiliation{M. N. Mikheev Institute of Metal Physics, Russian Academy of Sciences, 620108 Yekaterinburg, Russia}
\affiliation{Institute of Physics and Technology, Ural Federal University, 620002 Yekaterinburg, Russia}

\begin{abstract} 
We report a theoretical study of the effects of electron correlations and structural confinement on the electronic properties and magnetic state of LaNiO$_3$ (LNO) thin films epitaxially deposited on the $(001)$ LaAlO$_3$ (LAO) substrate. Using the DFT+U method
we compute the electronic band structure, magnetic properties, and phase stability of the 1.5 unit-cell-thick NiO$_2$-terminated LNO thin films. 
Our results reveal complex diversity of the electronic states caused by the effects of structural confinement, interfacial charge transfer and electronic correlations. Our calculations suggest the appearance of in-plane (110) charge disproportionation of the Ni ions in the interface NiO$_2$ layer of the antiferromagnetically ordered LNO thin films. Moreover, the electronic states of both the AFM and FM LNO/LAO show a large orbital polarization of the Ni ions in the surface NiO$_2$ layers. We propose the crucial importance of oxygen defects to explain the metal-to-insulator phase transition experimentally observed in a few-unit-cell-thick LNO/LAO thin films.
\end{abstract}

\maketitle

\section{Introduction}

The theoretical and experimental understanding of the electronic structure, magnetic state, and phase stability of
ultrathin films and heterostructures of strongly correlated transition metal oxides \cite{Imada,Khomskii} with perovskite crystal structure 
is of great challenge \cite{Catalan,Torrance,Garcia-Munoz,Boris,Hwang,Middey,King,Chen,Catalano,Golalikhani}. 
The atomic-scale layer-by-layer synthesis of these artificial materials with a nanoscale layer
thickness and controlled compositions makes it possible to design novel functional materials with
useful properties for a broad range of applications, e.g., in electronics, spintronics, energy storage, etc. \cite{Boris,Hwang,Middey,King,Chen,Catalano}. In these 
materials, the complex interplay between strong correlations, phase stability, and structural confinement results 
in diverse physical properties and rich phase diagrams \cite{Imada,Khomskii}. All these make such systems particularly interesting for fundamental 
research and applications \cite{Boris,Hwang,Middey,King,Chen,Catalano,Golalikhani}.

A particular interest has been devoted to the understanding of the electronic structure and magnetic properties of LaNiO$_3$ thin 
films (and heterostructures) epitaxially deposited on the LaAlO$_3$ or SrTiO$_3$ substrates (which are non-magnetic wide energy gap band insulators) \cite{Scherwitzl,Moon,Wu,Fowlie,Hepting,Ardizzone,
Guo_2020,Hansmann,Blanca-Romero,Doennig,Middey_2016,Geisler_2017,Geisler_2018,Geisler_2020,Geisler_2022,Lau_2013, Lau_2016,
Liau_2023,Lau,Georgescu}. In bulk LaNiO$_3$ (LNO) is a 
perovskite compound with rhombohedral symmetry ($R\bar{3}c$), the only member of the rare-earth nickelates 
$R$NiO$_3$ ($R$ - rare-earth element) that remains metallic at all temperatures due to the degenerate Ni $e_g$ orbitals (the Ni $t_{2g}$ orbitals are fully occupied) 
\cite{Catalan,Torrance,Garcia-Munoz,Ruppen,Bisogni, Guo}.
It has been shown that LNO is a negative charge-transfer materials with a nominal Ni $3d^8\underline{L}$ electronic configuration 
(with hole states at the O $2p$ orbitals). 
In contrast to LNO, all other representatives of bulk $R$NiO$_3$ (with $R$=Pr, Nd, Sm,.., Lu) exhibit a sharp metal-to-insulator transition (MIT) 
at low temperatures (and high pressure) \cite{Catalan,Torrance,Garcia-Munoz,Ruppen,Bisogni,Guo,Park_2012, Subedi, Seth, Hampel_2017, Peil, Hampel_2019, Haule, Liau_2021}. The MIT 
is accompanied by a structural transformation characterized by a cooperative breathing-mode distortion of NiO$_6$ octahedra and charge (bond) disproportionation of Ni ions \cite{Mazin,Johnston} with an alternating local-moment Ni$^{2+}$ $3d^8$ and nonmagnetic 
spin-singlet $3d^8\underline{L}^2$ electronic configurations in the low-temperature insulating phase. It is interesting to note that among bulk nickelates $R$NiO$_3$, in BiNiO$_3$ the MIT is associated with a cooperative breathing-mode 
distortion and charge disproportionation of the perovskite $A$-site Bi ions, while the Ni ions remain 2+ \cite{Azuma,Leonov_2019}. 
Moreover, charge (bond) disproportionation of Ni ions (charge-density wave instability) has also been recently proposed to occur in hole-doped infinite-layer superconducting nickelates $R$NiO$_2$ \cite{Li,Rossi,Tam,Krieger,Botana_2021,Slobodchikov,Leonov_2020,Kreisel,Karp,
Karp_2020,Lechermann,Leonov_2021}.

In LNO/LAO (LNO/STO) thin films structural confinement lifts the Ni $e_g$ orbital degeneracy both in the surface and at the interface layers, 
resulting in complex structural transformations (distortions of the perovskite lattice) with a large orbital polarization of Ni ions in the 
surface layer \cite{Boris,Hwang,Middey,King,Chen,Catalano,Golalikhani,
Scherwitzl,Moon,Wu,Fowlie,Hepting,Ardizzone,Guo_2020,Hansmann,Blanca-Romero,Doennig,Middey_2016,Geisler_2017,Geisler_2018,
Geisler_2020,Geisler_2022,Lau_2013, Lau_2016,Liau_2023,Lau,Georgescu}. 
Furthermore, the effects of structural confinement and strong correlations in the partially occupied Ni $3d$ shell in LNO thin films (and heterostructures) give rise to a metal-to-insulator phase transition
experimentally observed in such systems for a few-layer-thick LNO \cite{Boris,Hwang,Middey,King,Chen,Catalano,Golalikhani}. The MIT can be 
tuned by the LNO layer thickness and epitaxial strain. Moreover, it has been shown that
the electronic band structure and orbital polarization in LNO thin films are strongly impacted by the effects of surface termination and charge transfer.
The emergent complex electronic behavior enlarged with the possibility of fine-tuning of the electronic states and magnetic properties \cite{Vysotin,Hwang,Catalano} of such systems 
by applying epitaxial strain has attracted much recent attention \cite{Boris,Hwang,Middey,King,Chen,Catalano,Golalikhani,Scherwitzl,Moon,Wu,
Fowlie,Hepting,Ardizzone,Guo_2020,Hansmann,Blanca-Romero,Doennig,Middey_2016,Geisler_2017,Geisler_2018,Geisler_2020,
Geisler_2022,Lau_2013, Lau_2016,Liau_2023,Lau,Georgescu}. Nonetheless, the properties of LNO thin films are still poorly understood.

\section{Results and Discussion}

\begin{figure}[h]
\centering
\includegraphics[width=0.5\textwidth]{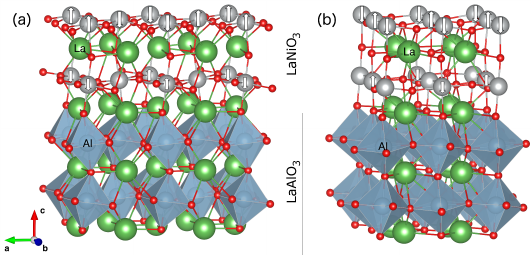}
\caption{
(Color online) Spin and charge disproportionated arrangements of the asymmetric (001) LNO/LAO slab obtained upon structural relaxation using the DFT+U method for the long-range $(\frac{1}{2}\frac{1}{2}\frac{1}{2})$ (a) and C-type AFM states (b). }
\label{Fig_1}
\end{figure}

\textbf{A. Computational details}. In our paper we study the electronic structure and magnetic properties of LaNiO$_3$ thin films epitaxially deposited on the (001) LaAlO$_3$ 
substrate. We explore the effects of electron correlations, structural confinement, and interfacial charge transfer on the electronic band 
structure and magnetic state of LNO/LAO using the DFT+U approach \cite{Anisimov,Dudarev}. 
%
In our calculations we adopt an asymmetric 
(001) LNO/LAO slab with the 1.5 unit-cell-thick NiO$_2$-terminated LNO thin film consisting of two atomic NiO$_2$ 
layers with a thickness of $\sim$5.75 \AA\ \cite{Boris,Hwang,Middey,King,Chen,Catalano,Golalikhani}. Note that Ni ions in the surface (top) layer 
are five-fold coordinated, with missing apical oxygens, 
while the Ni ions in the interfacial (bottom) NiO$_2$ layer have octahedral (six-fold) oxygen coordination. The LNO thin film is deposited 
on the (001) LAO substrate modelled by the six atomic-layer-thick LAO superlattice ($\sim$22.7 \AA) (see Fig.~\ref{Fig_1}). The LNO/LAO slab 
holds 8 f.u. LNO and 24 f.u. LAO, and is of 160 atoms in total. Each Ni-O (Al-O) layer contains four Ni (Al) ions. A vacuum layer width is 
taken of $\sim$16 \AA. 

We perform the DFT+U band structure calculations within generalized gradient approximation with Perdew-Burke-Ernzerhof exchange functional \cite{Perdew}
as implemented in the Quantum ESPRESSO package \cite{Baroni,Giannozzi}. In our DFT+U calculations we use ultrasoft pseudopotentials and the effective Coulomb 
interaction implemented within Dudarev's formulation \cite{Dudarev}, with $U_{eff}=U-J=5$~eV taken in accordance with previous estimates \cite{Blanca-Romero,Doennig,Middey_2016,Geisler_2017,Geisler_2018,Geisler_2020,Geisler_2022,Lau_2013, Lau_2016,Liau_2023,Lau,Georgescu}. We use a dipole 
moment correction as implemented in Ref.~\cite{Bengtsson} to retain a net dipole moment along the surface normal of the asymmetric LNO/LAO slab. 
In our calculations we consider a compressively strained (due to a lattice parameter mismatch) LAO substrate with the in-plane lattice 
parameter $a = 3.78$ \AA. Note that in bulk LNO the experimental in-plane lattice parameter is about 3.832 \AA\ at $T=100$ K \cite{Garcia-Munoz}. Obviously, the DFT+U calculations for the
tensile-strained substrates (e.g., for the SrTiO$_3$ or DyScO$_3$ substrates with the in-plane lattice constant $a=3.905$ \AA\ and 3.94 \AA, respectively) give 
enhanced correlation effects and, hence, stronger tendency towards localization \cite{Middey,Catalano,Blanca-Romero}. 

\begin{figure}[h]
\centering
\includegraphics[width=0.5\textwidth]{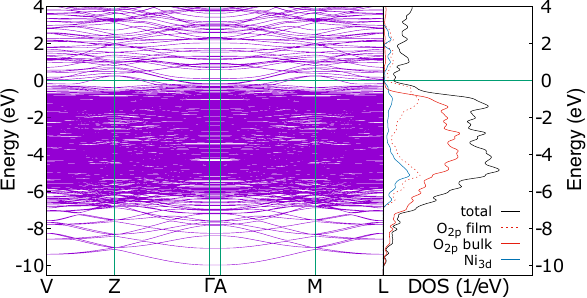}
\caption{
(Color online) Our DFT+U results for the band structure and orbitally-resolved spectral functions for the C-type AFM LNO/LAO. The top of the valence band is shown by dotted lines.}
\label{Fig_2}
\end{figure}

\begin{figure}[h]
\centering
\includegraphics[width=0.45\textwidth]{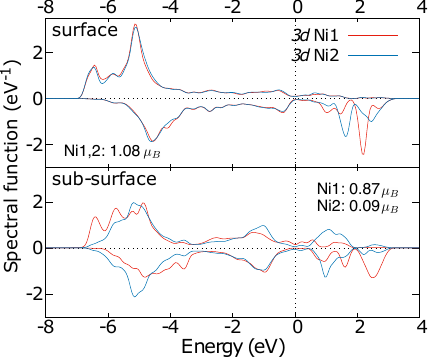}
\caption{
(Color online) Our DFT+U results for the Ni $3d$ spectral functions for the C-type AFM LNO/LAO. }
\label{Fig_3}
\end{figure}

\begin{figure}[h]
\includegraphics[width=0.5\textwidth]{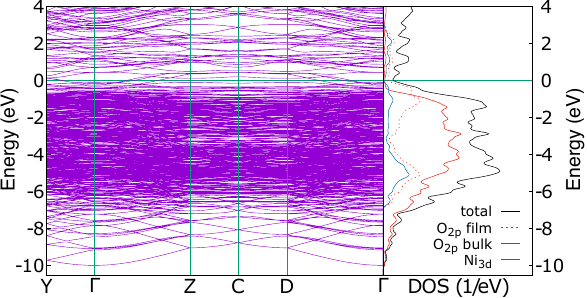}
\caption{
(Color online) Band structure and spectral functions for the $(\frac{1}{2}\frac{1}{2}\frac{1}{2})$ AFM LNO/LAO. }
\label{Fig_4}
\end{figure}

\begin{figure}[h]
\centering
\includegraphics[width=0.45\textwidth]{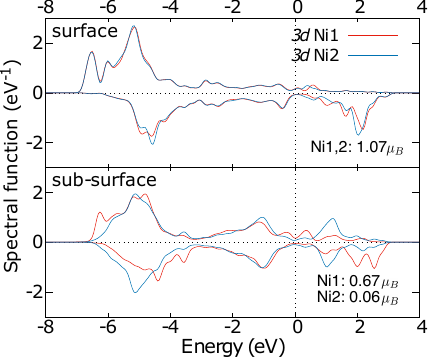}
\caption{
(Color online) Our results for the Ni $3d$ spectral functions for the $(\frac{1}{2}\frac{1}{2}\frac{1}{2})$ AFM LNO/LAO. }
\label{Fig_5}
\end{figure}

Using DFT+U we perform full structural optimization of the internal atomic positions of the 1.5 u.c. LNO film and the two-layer-thick interface 
LAO slab. The rest four-layer-thick LAO is considered as ``quasibulk'' with atomic positions fixed to that of the bulk LAO. Upon structural optimization 
the in-plane lattice constants and the angles between lattice vectors are constrained to that of LAO ($a = 3.78$ \AA). For the relaxed layers both tilting 
and rotation of NiO$_6$ (AlO$_6$) octahedra are allowed, as each Ni-O (Al-O) layers contain four Ni (Al) ions. 

In our study, we explore the interplay of long-range magnetic ordering and charge disproportionation of Ni ions tuned by structural confinement
on the properties of LNO/LAO thin films. In our calculations we consider 
different long-range magnetic arrangements. In particular, we adopt a long-range ferromagnetic (FM) and three different antiferromagnetic (AFM) 
arrangements, with a propagating wave vector ${\bf q}=(110)$ (C-type AFM), $(111)$ (G-type AFM ), and AFM $(\frac{1}{2}\frac{1}{2}\frac{1}{2})$, respectively. 
As a start, in our DFT+U calculations we set breathing-mode lattice distortions accompanied by charge disproportionation (charge ordering, 
CO) of the Ni sites with a wave vector (111). The same checkerboard (111) CO is observed in the low-temperature insulating phases of rare-earth 
nickelate perovskites $R$NiO$_3$ which is accompanied by the $(\frac{1}{2}\frac{1}{2}\frac{1}{2})$ AFM state \cite{Catalan,Torrance,Garcia-Munoz}. 

\textbf{B. AFM LNO/LAO}. In Fig.~\ref{Fig_1}  we show the spin and charge disproportionated states obtained upon structural optimization of the asymmetric $(001)$ LNO/LAO stab.
The DFT+U calculations yield a remarkable distortion of the Ni-O distances in LNO, with a significant deviation of the Ni-O bond 
lengths from that in the parent undistorted compound (with the average Ni-O bond length of 1.933 \AA). It leads to a tetragonal distortion of the crystal lattice. 
The crystal lattice is ``stretched'' along the $c$-axis by about 4 \%. Our results show a remarkable buckling of the surface layer, by about 0.3 \AA. 
For the C-type, G-type, and $(\frac{1}{2}\frac{1}{2}\frac{1}{2})$ AFM states we obtain metallic solution characterized by a cooperative breathing-mode distortion 
of the NiO$_6$ octahedra and charge disproportionation of Ni ions in the interface (bottom) NiO$_2$ layer. In fact, upon structural optimization
we obtain two distinct Ni sites characterized by two remarkably different average Ni-O bond lengths (with two structurally distinct Ni sites with ``expanded'' and ``compressed'' 
NiO$_6$ octahedra) and magnetic moments. We note that the average Ni-O bond lengths in the 
``expanded'' and ``compressed'' NiO$_6$ octahedra in the interface layer are $\sim$1.948-1.95 and 1.91-1.915 \AA, with a bond length difference 
of about 0.03-0.04 \AA. It is interesting to note that this value is compatible with an average bond length difference determined in other charge-disproportionated 
systems \cite{Attfield,Dalpian,Ivanova,Greenberg,Layek}. 
Our results for structural optimization show nearly regular octahedra with in-/out-of-plane Ni-O bond distances of 1.907/1.915 \AA\ for the ``compressed'' NiO$_6$ octahedra.
For the ``expanded'' NiO$_6$ octahedra a remarkable reduction of the out-of-plane bond distances with respect to the in-plane distances, about 1.934-to-1.955 \AA\ and 1.907-1.971 \AA\ for the C-type and for the $(\frac{1}{2}\frac{1}{2}\frac{1}{2})$ AFM is found, respectively. We also
note a large asymmetry of the out-of-plane Ni-O bond distances, about 0.1 \AA\ to the top and bottom apical oxygen ions associated with a remarkable shift of the
Ni site from the NiO$_6$ octahedron center (electric polarization active), by about 0.05 \AA .

In Figs.~\ref{Fig_2}-\ref{Fig_5} we summarize our results for the electronic band structure and orbitally-resolved spectral functions of LNA/LAO calculated by DFT+U for the C-type 
and $(\frac{1}{2}\frac{1}{2}\frac{1}{2})$ AFM.
Our DFT+U results show a relatively small (but robust) Ni $3d$ charge disproportionation of $\sim$0.1 for the Ni ions in the bottom NiO$_2$ layer \cite{Attfield,Dalpian,Slobodchikov}. 
Altogether, the total difference of the Ni $e_g$ occupation at the Ni$^{3-\delta}$ and Ni$^{3+\delta}$ sites is small, about 0.09 (implying a bond disproportionation scenario).
The obtained CO state has a characteristic wavevector $(110)$. Moreover, the DFT+U solution is characterized by two distinct Ni $3d$ magnetic 
moments of about 0.8 and 0.08 $\mu_B$ for the Ni ions with ``expanded'' and ``compressed'' 
oxygen coordination, respectively (all in the interface layer). This behavior is reminiscent of a site-selective Mott state \cite{Greenberg,Layek} observed in the low-temperature 
insulating phases of bulk rare-earth nickelates \cite{Park_2012,Liau_2021}. 
In addition, we observe strong hybridization between the Ni $3d$ and O $2p$ states, which results in a small spin polarization 
of nearby oxygen ions, with an opposite magnetic moment of $\sim$0.1 $\mu_B$. 

In contrast to the bottom layer, the total Ni $3d$ orbital occupations in the NiO$_2$ surface layer are nearly equal (homogeneous), 
with a rather small difference $\sim$0.02-0.04. The calculated Ni $3d$ magnetic moments are nearly the same, about 1.08 $\mu_B$ (for the AFM states). 
That is for the surface layer the DFT+U calculations give no evidence for the breathing-mode lattice distortions and charge disproportionation that makes the LNO/LAO interface metallic.
It is interesting to note that upon structural optimization of the initially breathing-mode distorted surface NiO$_2$ layer we obtain the lattice with
nearly equal average Ni-O bond lengths. This implies that charge disproportionation of Ni ions and cooperative breathing-mode distortions of the lattice are unstable 
in the surface NiO$_2$ layer. This is in contrast to the bottom layer where the breathing-mode lattice distortions and charge disproportionation are found to be robust. 
Our results for the in-/out-of-plane Ni-O distances in the surface layer NiO$_5$ square pyramids are 1.92/2.16-2.24 \AA.

We also notice the importance of interlayer charge transfer. Thus, the average total Ni $3d$ occupation in the surface layer is by $\sim$0.2 electron smaller than that in the 
interface NiO$_2$ layers. Moreover, for Ni ions in the surface layer we observe large spin-orbital polarization of about 0.58 with enhanced $3z^2-r^2$ minority spin hole density relative to 
the $x^2-y^2$ orbital states (driven be the $3d$ level splitting and strong correlation effects), implying the crucial importance of the quantum confinement and charge transfer effects. 
The total polarization of the Ni $e_g$ orbitals including both majority and minority spin channels is about 0.12.
Moreover, Ni ions in the interface NiO$_2$ layer reveal a rather weak
orbital polarization with an opposite sign (with an enhanced $x^2-y^2$ hole density), below $\sim$0.06. Our DFT+U total energy results suggest that the C-type AFM is energetically favorable compared to the AFM 
$(\frac{1}{2}\frac{1}{2}\frac{1}{2})$ solution, by $\sim$22 meV/Ni.
The total energy gain with respect to the G-type AFM is about 16~meV/Ni.

\begin{figure}[h]
\centering
\includegraphics[width=0.5\textwidth]{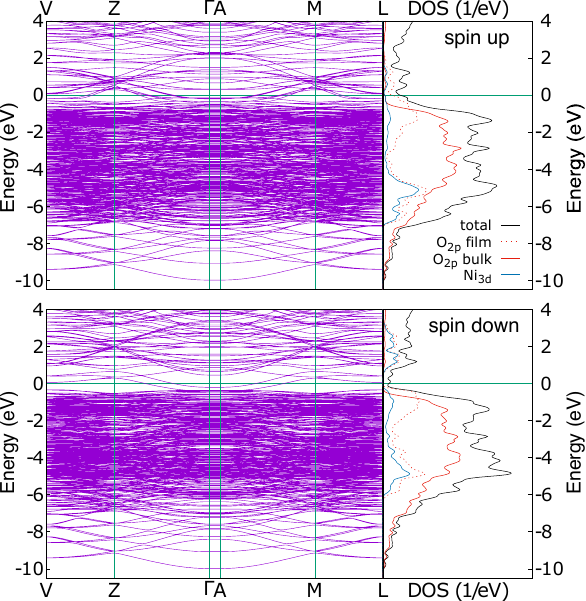}
\caption{
(Color online) Our results for the band structure dispersion and  spectral functions of FM LNO/LAO.}
\label{Fig_6}
\end{figure}

\begin{figure}[h]
\centering
\includegraphics[width=0.45\textwidth]{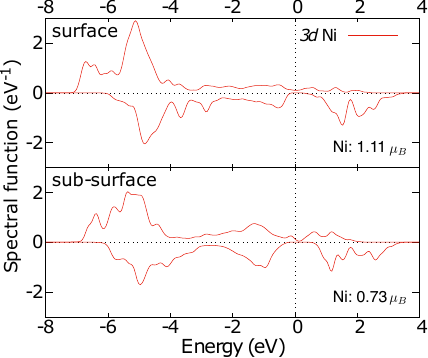}
\caption{
(Color online) Our results for the Ni $3d$ spectral functions obtained using DFT+U for the FM state.}
\label{Fig_7}
\end{figure}

\textbf{C. FM LNO/LAO}. In Figs.~\ref{Fig_6} and \ref{Fig_7} we show our DFT+U results for the band structure and orbitally-resolved spectral functions of LNA/LAO obtained for the FM state.
In contrast to the AFM LNO/LAO the calculations reveal no evidence for the in-plane charge disproportionation of Ni ions and breathing-mode distortions 
for both the surface and the interface NiO$_2$ layers (for the FM state). Upon structural optimization our DFT+U calculations yield nearly equal average 
Ni-O bond lengths for Ni ions for the surface ($\sim$1.967 \AA) 
and for the interfacial NiO$_2$ layers (1.937 \AA). The corresponding Ni-O in-/out-of-plane bond distances are 1.922/2.146 \AA\ for the surface and 1.936/1.938 \AA\ for the interface NiO$_2$ layers. In close similarity to the AMF results for 
the interface layer we notice a large asymmetry of the Ni-O bond distance between Ni ion and the top and bottom apical oxygens, about 1.897 and 1.979 \AA. That is Ni ion is shifted 
from the NiO$_6$ octahedron center by $\sim$0.04 \AA\ (electric polarization active). Moreover, for the surface layer the average Ni-O in-plane and out-of-plane bond distances are sufficiently different, 
about 1.922 and 2.146 \AA, respectively.

All the Ni ions in the NiO$_2$ surface layer are nearly identical.
The total Ni $3d$ orbital occupation for Ni ions in the NiO$_2$ surface layer is $\sim$8.56, while for the interface layer it is 8.74. The calculated Ni $3d$ magnetic 
moments are 1.11 and 0.73 $\mu_B$ for Ni ions in the surface and interface NiO$_2$ layers, respectively. Our results suggest that in-plane breathing-mode lattice distortion 
and charge disproportionation are unstable in FM LNA/LAO. However, we obtain a remarkable 
interlayer charge transfer (between the Ni ions in the surface and interfacial NiO$_2$ layers), implying the importance of structural confinement effects. In fact, the average 
total Ni $3d$ occupation in the surface layer is by $\sim$0.18 electron smaller than that in the interface NiO$_2$ layers. Furthermore, for the Ni ions in the surface layer our 
DFT+U results give large spin-orbital polarization of about 0.51, with nearly empty $3z^2-r^2$ minority spin orbitals (its orbital occupation 0.18), relative to the $x^2-y^2$ states (0.56). 
The total orbital polarization including the spin majority and minority Ni $e_g$ states is considerably smaller, about 0.19. This highlights the 
remarkable interplay of the quantum confinement and charge transfer effects in LNA/LAO. Moreover, Ni ions in the bottom layer show no sizeable orbital polarization 
effects, below $\sim$0.03, with an enhanced $x^2-y^2$ hole density. Our total energy calculations show that the FM LNO/LAO is energetically favorable compared to the C-type AFM, 
by $\sim$29 meV/Ni, suggesting that LNO/LAO
interface is ferromagnetic. 
Similar to AFM LNO/LAO we observe strong hybridization between the Ni $3d$ and O $2p$ states, which results in a small spin polarization 
of nearby oxygen ions, with an opposite magnetic moment of $\sim$0.1-0.2 $\mu_B$.

Our DFT+U calculations show that structural confinement and relaxations effects in combination with strong correlations alone are not sufficient to explain the 
experimentally observed insulating state of a 1.5 unit-cell-thick LNO/LAO, implying the possible importance of oxygen defects and non-local correlations, in agreement with Refs.~\cite {Liau_2021,Golalikhani}. 
In fact, our DFT+U calculations for the FM, C-type, G-type, and $(\frac{1}{2}\frac{1}{2}\frac{1}{2})$ AFM LNO/LAO exhibit small but still nonzero spectral weight at the Fermi energy. 
The DFT+U calculations yield a pseudo-gap state with a broad parabolic-like band at the Fermi level. Its origin is related to the strongly hybridizing Ni $e_g$ and O $2p$ orbitals, 
crossing the $E_F$ near the Brillouin zone (BZ) $\Gamma$ in the $(\frac{1}{2}\frac{1}{2}\frac{1}{2})$ AFM and near the BZ  $\Gamma$ and L points in the C-type and G-type AFM and FM states, implying the importance of negative charge-transfer effects at the 
surface NiO$_2$ layer. 
We therefore propose that the nature of insulating state of a few-unit-cell-thick LNO/LAO films may be reminiscent the mechanism of the metal-insulator transition in oxygen-deficient nickelate
LaNiO$_{3-\delta}$ \cite{Liau_2023,Golalikhani}.

\section{Conclusion}

In conclusion, using the DFT+U method we computed the electronic and magnetic properties of LaNiO$_3$ thin films epitaxially deposited on 
the $(001)$ LaAlO$_3$ substrate. Our results reveal a remarkable interplay of the effects of electron correlations, structural confinement, and interfacial charge transfer on the electronic band 
structure and magnetic state of the LNO thin films. 
It was shown that the in-plane breathing-mode lattice distortion and charge disproportionation of Ni ions in the surface NO$_2$ layer are unstable. The surface layer is metallic, charge-homogeneous, although with a large orbital polarization of the Ni ions. Similar behavior we observe for the long-range FM ordered LNO/LAO thin films. Irrespective to the magnetic state we observe a sufficient charge disproportionation between the Ni ions in the surface and interfacial NiO$_2$ layers, implying the importance of interlayer charge transfer and strong correlations.
Our DFT+U results show that structural confinement and relaxations effects in combination with strong correlations alone are not sufficient to explain the 
experimentally observed insulating state of a 1.5 unit-cell-thick LNO/LAO. This suggests the possible importance of oxygen defects to explain a metal-to-insulator phase transition experimentally observed in a few-unit-cell-thick LNO/LAO thin films. We believe that this topic deserves further detailed theoretical and experimental considerations.

We acknowledge support by the Russian Science Foundation Project No. 22-22-00926, https://rscf.ru/project/22-22-00926.


\end{document}